\documentclass[11pt]{article}
\usepackage{cite}
\usepackage{amsmath,amssymb,amsfonts}
\usepackage{algorithm}
\usepackage{algpseudocode}
\usepackage{graphicx}
\usepackage{subfigure}
\usepackage{textcomp}
\usepackage{bm}
\usepackage{booktabs}
\usepackage{multirow}
\usepackage{color}
\usepackage[a4paper]{geometry}
\usepackage{setspace}
\usepackage{makecell}
\usepackage{url}
\usepackage{authblk}
\usepackage{caption}

\renewcommand\footnotemark{}

\begin{document}
\setlength{\belowdisplayskip}{10pt} \setlength{\abovedisplayskip}{10pt}
\captionsetup{font={small}}

\title{Physics-/Model-Based and Data-Driven Methods for Low-Dose Computed Tomography: A survey}
\author{Wenjun Xia, \textit{Graduate Student Member, IEEE}, Hongming Shan, \textit{Senior Member, IEEE}, Ge Wang, \textit{Fellow, IEEE}, and Yi Zhang, \textit{Senior Member, IEEE}
\thanks{
This work was supported in part by the Sichuan Science and
Technology Program under Grant 2021JDJQ0024, in part by the Sichuan University ’From 0 to 1’ Innovative Research Program under Grant 2022SCUH0016, in part by the National Natural Science Foundation of China under Grant 62101136, in part by the Shanghai Sailing Program under Grant 21YF1402800, in part by the Shanghai Municipal of Science and Technology Project under Grant 20JC1419500, in part by Shanghai Center for Brain Science and Brain-inspired Technology, in part by the National Institute of Biomedical Imaging and Bioengineering (NIBIB) of the National Institutes of Health (NIH) under Grant R01EB026646 , and in part by the National Institute of General Medical Sciences (NIGMS) of the National Institutes of Health (NIH) under Grant R42GM142394 (Corresponding author: Y. Zhang).

W. Xia and Y. Zhang are with the School of Cyber Science and Engineering, Sichuan University, Chengdu 610065, China (e-mail: xwj90620@gmail.com; yzhang@scu.edu.cn).

H. Shan is with the Institute of Science and Technology for Brain-Inspired Intelligence and MOE Frontiers Center for Brain Science, Fudan University, Shanghai, and also with Shanghai Center for Brain Science and Brain-Inspired Technology, Shanghai 200433, China (email: hmshan@fudan.edu.cn).

G. Wang is with the Department of Biomedical Engineering, Rensselaer Polytechnic Institute, Troy, NY 12180 USA (email: wangg6@rpi.edu).}}

\date{}
\maketitle
\vspace*{-1.0cm}
\begin{abstract}
Since 2016, deep learning (DL) has advanced tomographic imaging with remarkable successes, especially in low-dose computed tomography (LDCT) imaging.
Despite being driven by big data, the LDCT denoising and pure end-to-end reconstruction networks often suffer from the black box nature and major issues such as instabilities, which is a major barrier to apply deep learning methods in low-dose CT applications. 
An emerging trend is to integrate imaging physics and model into deep networks, enabling a hybridization of physics/model-based and data-driven elements. 
In this paper, we systematically review the physics/model-based data-driven methods for LDCT, summarize the loss functions and training strategies, evaluate the performance of different methods, and discuss relevant issues and future directions. 
\end{abstract}

\section{Introduction}
\label{sec:introduction}
Since the invention of computed tomography (CT) in 1970s, it has become an indispensable imaging modality for screening, diagnosis, and therapeutic planning.
Due to the potential damage to healthy tissues, the radiation dose minimization for x-ray CT has been widely studied over past two decades.
In some major clinical tasks, the radiation dose of a single CT scan can be up to 43 milli-Sieverts (mSv)~\cite{smith2009radiation}, which is an order of magnitude higher than the amount of the natural background radiation one receives annually.
The radiation dose can be reduced by lowering the x-ray flux physically, 
which is called low-dose CT (LDCT).
However, LDCT will degrade the signal-to-noise ratio (SNR) and compromise the subsequent image quality.

The conventional tomographic reconstruction algorithm can hardly achieve satisfactory LDCT image quality.
To meet the clinical requirements, advanced algorithms are required to suppress noise and artifacts associated with LDCT. 
Up to now, promising results have been obtained, improving LDCT quality and diagnostic performance in various clinical scenarios.
Generally speaking, LDCT algorithms can be divided into four categories: sinogram domain filtering, image domain post-processing, model-based iterative reconstruction (MBIR), and deep learning (DL) methods.

Sinogram domain filtering directly performs denoising in the space of projection data. Then, the denoised raw data can be reconstructed into high-quality CT images using analytic algorithms. 
Depending on the noise distribution, appropriate filters can be designed.
Structural adaptive filtering~\cite{balda2012ray} 
is a representative algorithm in this category, which effectively refines the clarity of LDCT images. 
The main advantage of sinogram domain filtering is that it can suppress noise based on the known distribution. 
However, any model mismatch or inappropriate operations in the projection domain will introduce global interference, compromising the accuracy and robustness of sinogram domain filtering results.

Image domain post-processing is more flexible and stable than sinogram domain filtering.
Based on appropriate prior assumptions of CT image, such as sparsity, 
several popular methods were developed~\cite{chen2013improving}.
These methods can effectively denoise LDCT images, but their prototypes were often developed for natural image processing. 
In many aspects, the properties of LDCT are quite different from natural images. 
For example, LDCT image noise does not follow any known distribution, depends on underlying structures, and is difficult to model analytically. The image noise distribution is complex, and so is the image content prior. These are responsible for limited performance of image domain post-processing.

MBIR combines the advantages of the two kinds of methods mentioned above and works to minimize an energy-based objective function.
The energy model usually consists of two parts: the fidelity term with the noise model in the projection domain and the regularization term with the prior model in the image domain.
Since the noise model for LDCT in the projection domain is well-established, research efforts in developing MBIR are more focused on the prior model.
Utilizing the well-known image sparsity for LDCT, a number of methods were proposed~\cite{yu2005total, wen2015structured, chun2017sparse}.
The MBIR algorithms usually deliver robust performance and achieve clinically satisfactory results after the regularization terms are properly designed along with well-tuned balancing parameters. 
However, these requirements for an MBIR algorithm may restrict its applicability. Customizing an MBIR algorithm takes extensive experience and skills. Also, MBIR algorithms suffer from expensively computational cost.

Recently, DL was introduced for tomographic imaging.
Driven by big data, DL promises to overcome the main shortcoming of conventional algorithms which demands the explicit design of regularizers and cannot guarantee the optimality and generalizability.
The DL methods extract information from a large amount of data to ensure the objectivity and comprehensiveness of the extracted information.
By learning the mapping from LDCT scans to normal-dose CT (NDCT) images, a series of studies were performed~\cite{chen2017lowdose, yang2018low,kang2019cycle,shan2019competitive, nishii2022deep}.
These methods can be seen as a combination of image domain post-processing and data-driven methods.
They inherit the advantages of the post-processing algorithms and DL methods, and have high processing efficiency, excellent performance, and great clinical potential.
However, they also have drawbacks.
These methods usually use the approximate or pseudo-inversion of the raw data as the input of the network.
The initially reconstructed images may miss some structures, which cannot be easily restored by the network if raw data are unavailable.
On the other hand, noise and artifacts in filtered back-projection (FBP) reconstructions could be perceived as meaningful structures by a denoising network.
Both circumstances will compromise the diagnostic performance, resulting in either false positives or false negatives.

\begin{figure}[tbp]
	\centering
	\includegraphics[width=0.4\textwidth]{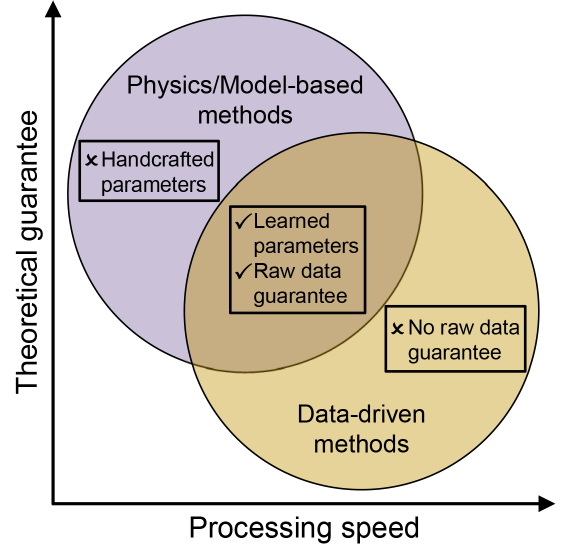}
	\caption{Advantages of synergizing physics/model-based methods and data-driven methods.}
	\label{fig:1}
\end{figure}

Naturally, synergizing physics/model-based methods and data-driven methods will enjoy the best of both worlds.
While deep image denoising only handles reconstructed images, MBIR methods are more robust and safer. In each iteration of MBIR, raw data will be used to rectify intermediate results and improve the data consistency.
By introducing the CT physics or MBIR model, researchers can embed the raw data constraint into the network, which avoids the information loss in the process of image reconstruction.
Over the past years, a number of physics/model-based data-driven methods for LDCT have been proposed~\cite{gupta2018cnn,chen2018learn,adler2018learned}.
As shown in Fig.~\ref{fig:1}, these methods address the shortcomings of the physics/model-based methods and data-driven networks, and achieve an excellent balance between the improved accuracy with learned parameters and the robustness aided by data fidelity.

In this paper, we will review these methods. In the next section, the problem of LDCT is described, and the conventional modeling and optimization methods are introduced. In the third section, different kinds of methods are summarized to incorporate physics/model into deep learning framework. In the fourth section, several experiments are conducted to compare different hybrid methods for LDCT. In the fifth section, we discuss relevant issues. Finally, we will conclude the paper in the last section.

\section{Physics/Model-based LDCT Methods}
\label{sec:physics}
\subsection{CT Physics}
Assuming that an x-ray tube has an incident flux $I_0$ which can be measured in an air scan, the number of photons received by a detector, $I$, can be formulated as
$
	I = I_0 \exp\left(-\int_{l} \mu \, \mathrm{d}l\right)
$, where $\mu$ is the linear attenuation coefficient, and $l$ represents the x-ray path.
After a logarithmic transformation, the line integral can be obtained as
\begin{align}
	-\log \frac{I}{I_0} = \int_{l} \mu \, \mathrm{d}l.
	\label{eq:2}
\end{align}
Such line integrals are typically organized into projections and stored as a sinogram.
The line integrals in the form of Eq.~\eqref{eq:2} can be discretized into a linear system 
$
	\bm{y} = A \bm{x}
$, where $\bm{x} \in \mathbb{R}^N$ denotes the attenuation coefficient distribution to be solved, 
$\bm{y} \in \mathbb{R}^M$ represents the projection data, and $A \in \mathbb{R}^{M \times N}$ is the system matrix for a pre-specified scanning geometry.

\subsection{CT Noise}
The noise in CT data  mainly consists of the following two components~\cite{diwakar2018review}:
\begin{enumerate}

\item[1)] \textbf{Statistical noise:}
Statistical noise, also known as quantum noise, is the main noise component in LDCT and originates from statistical fluctuations in the emission of x-ray photons.

\item[2)] \textbf{Electronic noise:}
Electronic noise occurs when analog signals are converted into digital signals. 

\end{enumerate}

\subsection{Simulation}
In clinical practice, it is difficult to obtain paired LDCT and NDCT datasets from two separate scans due to uncontrollable organ movement and radiation dose limitation. As a result, numerical simulation is important to produce LDCT data from an NDCT scan.

In~\cite{yu2012development}, a noise simulation method was proposed for LDCT research, and applied to generate the public dataset ``\textit{the
2016 NIH-AAPM-Mayo Clinic Low-Dose CT Grand Challenge}".
The number of detected x-ray photons can be approximately considered as normally distributed, and formulated as: 
$
	\bm{\tilde{y}} = \bm{y} + \sqrt{\frac{1-a}{a}\cdot \frac{\exp(\bm{y})}{I_0}\cdot \left(1+\frac{1+a}{a}\cdot \frac{\sigma_e^2 \cdot \exp(\bm{y})}{I_0}\right)}\cdot \xi,
$
where $\xi \sim \mathcal{N}(0,1)$, $\sigma_e^2$ represents the variance of electronic noise, and $a$ denotes the dose factor.

\subsection{Conventional CT Reconstruction Methods}
Conventional CT image reconstruction takes both measurement data and prior knowledge into account, and is performed by minimizing an energy model iteratively. The general energy model for LDCT reconstruction can be formulated as
\begin{equation}
	\min_{\bm{x}} \ \Phi(\bm{x}) + \lambda R(\bm{x}),
	\label{eq:general_energy_model}
\end{equation}
which has two parts: a fidelity term $\Phi(\bm{x})$ and a regularization term $R(\bm{x})$, with $\lambda$ being the penalty parameter.

The fidelity term is a metric of the reconstruction result measuring the consistence to the measurement data. The weighted least-squares (WLS) function is usually adopted as the fidelity term:
$
	\Phi(\bm{x}) = \frac{1}{2} \left\| \bm{\tilde{y}}-A\bm{x} \right\|_{\Sigma^{-1}}^2 = \frac{1}{2} (\bm{\tilde{y}}-A\bm{x})^T \Sigma^{-1} (\bm{\tilde{y}}-A\bm{x}),
	\label{eq9}
$
where $\Sigma$ is a diagonal matrix with its elements on the main diagonal $\Sigma_{ii}=\sigma_i^2$ being the estimated variances of data. Since the ideal flux is unknown, the variance is usually estimated as $\sigma_i^2 = (\tilde{I}_i+\sigma_e^2)/\tilde{I}_i^2$ where $\tilde{I}_i=a I_{0}\exp(-\bm{\tilde{y}}_i)$~\cite{chun2017sparse}. 

Dedicated regularization terms were designed for different types of images, depending on the nature of images and researchers' expertise.
Over the past years, various regularization terms were proposed along with the improved understanding of CT image properties.
Importantly, the well-known sparsity can be expressed as
$
	R(\bm{x}) = \left\| W\bm{x} \right\|_1,
$
where $\left\|\cdot \right\|_1$ is the $\ell_1$ norm, and $W$ is a sparsifying transform matrix.
Commonly used sparsifying transforms include the gradient transform (total variation)~\cite{yu2005total},
learned sparsifying transform~\cite{wen2015structured, chun2017sparse, zheng2018pwls}, 
etc.
Subsequently, further leveraging the two-dimensional structure of an image, low-rank became popular for LDCT reconstruction~\cite{cai2014cine, kim2014sparse, xia2019spectral, chen2022font}.
The low-rank constraint can be relaxed to minimization of the nuclear norm
$
	R(\bm{x}) = \left\| \bm{x} \right\|_*=\sum_i \sigma_i(\bm{x}),
$
where $\sigma_i(\bm{x})$ is the $i$-th largest singular value of $\bm{x}$.

Generally, these models in Eq.~\eqref{eq:general_energy_model} do not have closed solutions and need to be iteratively optimized.
Sometimes, auxiliary and dual variables are introduced to simplify the calculation and facilitate the convergence. 
The idea of introducing auxiliary variables is in the same spirit of plug-and-play (PnP)~\cite{venkatakrishnan2013plug, rond2016poisson}, which can decouple the primal problem and inverse conveniently.
For example, based on the PnP scheme with the WLS fidelity, Eq.~\eqref{eq:general_energy_model} can be rewritten as
\begin{equation}
	\min_{\bm{x},\bm{v}} \ \frac{1}{2} (\bm{\tilde{y}}-A\bm{x})^T \Sigma^{-1} (\bm{\tilde{y}}-A\bm{x}) + \lambda R(\bm{v}), \ \mathrm{s.t.} \ \bm{x}=\bm{v}.
	\label{eq:12}
\end{equation}
Then, the primal variable $\bm{x}$ and auxiliary variable $\bm{v}$ can be alternately optimized.
Two representative alternating optimization algorithms are ADMM and Split-Bregman, which divide the model into sub-problems and solve them accordingly.

\section{Physics/Model-based Data-driven Methods}
\label{sec:physics_data}
The popular approach for LDCT denoising with DL employs convolution layers and activation functions to build a neural network,
whose input and output are both images.
These methods are simple to implement and deliver impressive denoising performance, but they can hardly recover details lost in the input image.
On the other hand, the MBIR algorithm is safer.
In each iteration of MBIR, it uses the measurement to correct an intermediate result. Constrained by the measurement, the MBIR result respects the data consistency and restores missing structures well in the reconstructed image.
Following the idea of the DL-based post-processing method,
it is natural to synergize the physics/model-based and data-driven methods.
Such hybrid methods would not only have data-driven benefits but also have better robustness and interpretability out of the physics/model-based formulation.
Table~\ref{summary_1} summarizes these methods.
The rest of this section introduces this kind of method.

\begin{figure*}[tbp]
	\centering
	\includegraphics[width=1.0\textwidth]{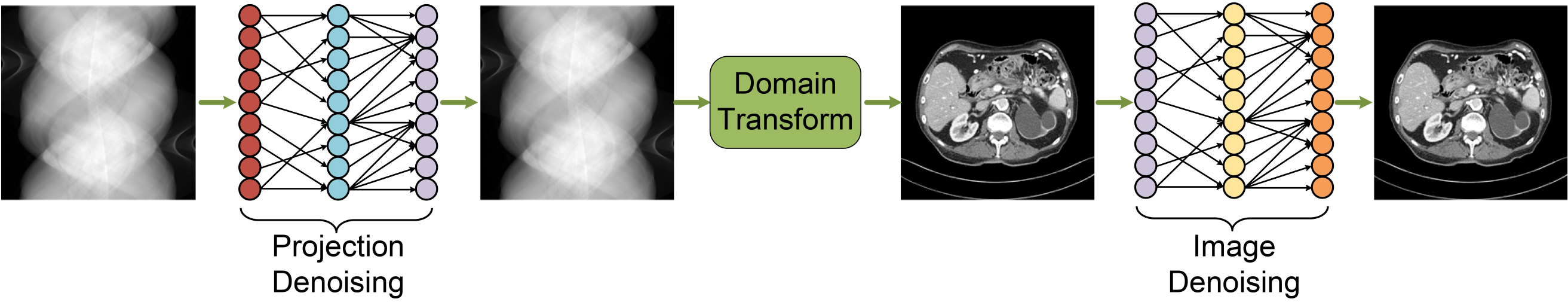}
	\caption{General workflow for the physics-based data-driven LDCT denoising.}
	\label{fig:3}
\end{figure*}

\subsection{Physics-based Data-driven Methods}
As shown in Fig.~\ref{fig:3}, the physics-based data-driven methods include a differentiable domain transform based on the CT physics between the projection and image domains in the network. The input and output of the network are usually projection data and image data, respectively. In the beginning, the network used a conventional domain transform from the projection domain to the image domain. Then, inspired by the work on learned domain transformation, researchers built networks with fully connected (FC) layers to replace the conventional domain transform and learn the inverse Radon transform directly~\cite{he2020radon}.

\begin{table}[tbp]
\setlength\tabcolsep{5pt}
\small
\centering
\caption{Representative physics/model-based data-driven methods}
\label{summary_1}
\begin{tabular}{cll}
\toprule
&\multicolumn{1}{c}{Reference}&\multicolumn{1}{c}{Highlight} \\
\midrule
-&W{\"u}rfl \textit{et al.}~\cite{wurfl2018deep} & Learned filters of FBP \\ 
HDNet&Hu \textit{et al.}~\cite{hu2020hybrid}   & Dual domain processing with FBP\\ 

VVBPTensorNet&Tao \textit{et al.}~\cite{tao2021learning}   & Filtered back-projection view-by-view \\ 
CLEAR& Zhang \textit{et al.}~\cite{zhang2021clear} & Multi-level consistency loss \\
\midrule
AUTOMAP&Zhu \textit{et al.}~\cite{zhu2018image}   & Learned transform with FC\\
iRadonMap&He \textit{et al.}~\cite{he2020radon}  & Transform based on scan trajectory\\ 
DSigNet&He \textit{et al.}~\cite{he2021downsampled}   & Downsampling of geometry and volume\\ 
iCTNet&Li \textit{et al.}~\cite{li2019learning}    & Shared parameters for different views\\
-&Fu \textit{et al.}~\cite{fu2019hierarchical}  & Hierarchical architecture\\
\midrule
KSAE&Wu \textit{et al.}~\cite{wu2017iterative} & Learned sparsifying transform\\
CNNRPGD&Gupta \textit{et al.}~\cite{gupta2018cnn} & Learned projection operation\\
REDAEP&Zhang \textit{et al.}~\cite{zhang2020redaep} & Learned denoising autoencoding prior\\
MomentumNet&Chun \textit{et al.}~\cite{chun2020momentum}  & Momentum-based extrapolation\\
SUPER&Ye \textit{et al.}~\cite{ye2021unified} & Combination of learned regularizations\\
\midrule
LEARN&Chen \textit{et al.}~\cite{chen2018learn}  & Unrolled gradient descent algorithm\\
LPD&Adler \textit{et al.}~\cite{adler2018learned} & Unrolled PDHG algorithm \\
AHP-Net&Ding \textit{et al.}~\cite{ding2021deep}  & Learned hyperparameters with FC\\
MetaInvNet&Zhang \textit{et al.}~\cite{zhang2020metainv} & Learned initializer for conjugate gradient\\
FistaNet&Xiang \textit{et al.}~\cite{xiang2021fista} & Unrolled FISTA algorithm\\
\bottomrule
\end{tabular}
\end{table}

\paragraph{Conventional transform}
The architecture of this kind of network is often featured by two sub-networks: one in the projection domain and the other in the image domain, both of which are usually implemented with convolutional neural networks (CNNs). The projection data is fed into the first sub-network, in which the measurement data can be denoised. The denoised projection is then converted into an image using a differentiable conventional transformation. Finally, the image is processed by the second sub-network to improve the reconstruction quality.
Many network architectures widely used for image processing in the literature can be adapted into the sub-networks in both domains. 
Since the statistical distributions of CT noise in the projection and  image domains are quite different, the combination of the denoising processes in the two domains can be complementary, making the denoising process more effective and more stable.
The differentiable domain transform allows the information exchange between the two sub-networks.
The simplest domain transform is back-projection, which is a differentiable linear transform~\cite{wurfl2018deep}.
A more reasonable yet very efficient transform is filtered back-projection (FBP)~\cite{hu2020hybrid, zhang2021clear}. 
A main advantage of FBP lies in that the projection data can be directly transformed into a suitable numerical range, which is more friendly to the subsequent image domain processing.
Another interesting domain transform is the FBP view-by-view~\cite{tao2021learning}.
This transform back-projects the projection data into multi-channels in the image space, each of which is the back-projection from one projection view. 
It decouples the data from multi-views to obtain more information.
These domain transforms are limited by the understanding and modeling of CT physics.
With deep learning, it is feasible to learn the involved kernels and perform the domain transform.

\begin{algorithm}[t]  
\small
	\caption{Training a denoiser-based method in an iteration-independent/dependent fashion.}
	\label{alg:denoiser}  
	\begin{algorithmic}
		\Require Training set $\{\bm{\tilde{y}}_i, \bm{\hat{x}}_i\}_{i=1}^{N_s}$, Denoiser $D_{\theta}$
		\Ensure $\bm{x}^0_i = \mathrm{FBP}(\bm{\tilde{y}}_i), i = 1,2,\dots,N_s$
		\\\hrulefill
		\State \textbf{Iteration-independent denoiser:}
		\State \textbf{Train $D_{\theta}$:} $\theta = \arg\min_{\theta} \frac{1}{N_s} \sum_{i=1}^{N_s} \|D_{\theta}(\bm{x}^0_i) - \bm{\hat{x}}_i\|_2^2$
		\For{$t=0,1,...,N_t-1$}
		\State \textbf{Obtain} $\bm{x}^{t+1}$ from $D_{\theta}(\bm{x}^t)$
		\EndFor \\
		\Return $\bm{x}^{N_t}$
		\\\hrulefill
		\State \textbf{Iteration-dependent denoiser:}
		\For{$t=0,1,...,N_t-1$}
		\State \textbf{Train $D_{\theta}$:} $\theta^t = \arg\min_{\theta} \frac{1}{N_s} \sum_{i=1}^{N_s} \|D_{\theta}(\bm{x}^t_i) - \bm{\hat{x}}_i\|_2^2$
		\State \textbf{Obtain} $\bm{x}^{t+1}$ from $D_{\theta^t}(\bm{x}^t)$
		\EndFor \\
		\Return $\bm{x}^{N_t}$
	\end{algorithmic}  
\end{algorithm}

\paragraph{Learned transform}
The learned transform can use FC layers to learn the physics-based transform from the projection domain to the image space. 
AUTOMAP is a representative network, which maps tomographic data to a reconstructed image through FC layers~\cite{zhu2018image}.
However, such an architecture would be unaffordable in most cases of medical images because of the expensive computational and memory costs.
As a result, major efforts were made to improve the learned transforms by reducing the computational overhead~\cite{he2020radon, he2021downsampled, li2019learning}.
Since each pixel traces a sinusoidal curve in the projection domain,
\cite{he2020radon} proposed to sum linearly along the trajectory so that the weights of FC layers are sparse.
In an improved version of this work, the geometry and volume were down-sampled to further reduce the computational cost~\cite{he2021downsampled}.
Another effective way to reduce the cost is to use shared parameters.
In~\cite{li2019learning}, the measurements of different views are processed with shared parameters for the domain transform.
And in~\cite{fu2019hierarchical}, authors proposed a hierarchical architecture, where the shared parameters are gradually localized to the pixel level.
Compared with the conventional transform, the learned transform has the potential to achieve better performance. 

By incorporating the CT physics into the denoising process and working in both projection and image domains, image denoising can be effectively performed. However, the issue of generalizability is important for clinical applications.
The learned transform has limited generalizability because it can only be applied for a fixed imaging geometry.
When the geometry and volume differ from what is assumed in the training setup, the trained network will be inapplicable.
In contrast, the traditional transform is more stable and only needs to adjust the corresponding parameters for different geometries and volumes.
Therefore, the further development of learned transforms needs to make them more flexible and more generalizable.

\subsection{Model-based Data-driven Methods}
Given the generalizability, stability and interpretability of the MBIR algorithm, it is desirable to combine MBIR and DL for LDCT denoising. Deep learning is effective in solving complicated problems with big data. MBIR-based reconstruction has a fixed fidelity term and needs efforts to find a good regularizer. For model-based data-driven reconstruction, researchers replaced the handcrafted regularization terms with neural networks and produced results often superior to the traditional MBIR counterparts.
By the way of embedding a neural network into the MBIR scheme, we can divide the model-based data-driven methods into two categories: denoiser and unrolling.

\paragraph{Denoiser}
This approach follows the conventional iterative optimization scheme.
In its iterative process, a neural network is introduced with the idea of PnP~\cite{rond2016poisson}.
Drawing on the framework of PnP, a regularization by denoising (RED) was proposed~\cite{romano2017little}.
With this regularization, the CT optimization model can be formulated as
\begin{equation}
	\min_{\bm{x}} \ \frac{1}{2} (\bm{\tilde{y}}-A\bm{x})^T \Sigma^{-1} (\bm{\tilde{y}}-A\bm{x}) + \frac{\lambda}{2} \bm{x}^T(\bm{x}-D(\bm{x})),
	\label{eq:14}
\end{equation}
where $D(\cdot)$ is a denoiser implemented by the neural network.
The optimization process can be expressed as
\begin{equation}
	\begin{split}
	\bm{x}^{t+\frac{1}{2}} &= \arg\min_{\bm{x}} \ \frac{1}{2} (\bm{\tilde{y}}-A\bm{x})^T \Sigma^{-1} (\bm{\tilde{y}}-A\bm{x}) + \frac{\alpha}{2} \|\bm{x} - \bm{x}^{t}\|_2^2, \\
	\bm{x}^t &= \arg\min_{\bm{x}} \ \frac{\beta}{2} \|\bm{x} - \bm{x}^{t+\frac{1}{2}}\|_2^2 + \frac{\lambda}{2} \bm{x}^T(\bm{x}-D(\bm{x}^{t+\frac{1}{2}})).
	\end{split}
\label{eq:15}
\end{equation}
The optimization of $\bm{x}^{t+\frac{1}{2}}$ can be done using a method for solving the quadratic problem.
The solution of $\bm{x}^t$ can be obtained either directly from  the denoiser
\begin{equation}
	\bm{x}^t = D(\bm{x}^{t+\frac{1}{2}}),
	\label{eq:16}
\end{equation}
or as a semi-denoised result
\begin{equation}
	\bm{x}^t = \left(1-\frac{\lambda}{\beta+\lambda}\right) \bm{x}^{t+\frac{1}{2}} + \frac{\lambda}{\beta+\lambda} D(\bm{x}^{t+\frac{1}{2}}).
	\label{eq:17}
\end{equation}
Of course, there are other solutions and combinations~\cite{ye2021unified, zhang2020redaep}.

There are two ways to train the denoiser. The first is to train a general denoiser for all iterations~\cite{wu2017iterative, gupta2018cnn, zhang2020redaep}.
In this option, the denoiser can be obtained with noisy images and the corresponding labels as training pairs.
However, it is difficult for the denoiser to achieve optimal denoising performance in each iteration.
The second option can partially solve this problem by training an iteration-dependent denoiser dynamically~\cite{chun2020momentum, ye2021unified}.
In each iteration, the denoiser will denoise an intermediate image with different parameters to  optimize denoising performance.
Of course, training model parameters for each iteration will demand a much higher computational cost.
\textit{Algorithm~\ref{alg:denoiser}} supports either iteration-independent or iteration-dependent denoisers, where
$\bm{\tilde{y}}$ denotes projection data, $\bm{\hat{x}}$ represents a noise-free image, and the mean square error (MSE) is assumed as the loss function.

\begin{figure*}[tbp]
	\centering
	\includegraphics[width=1.0\textwidth]{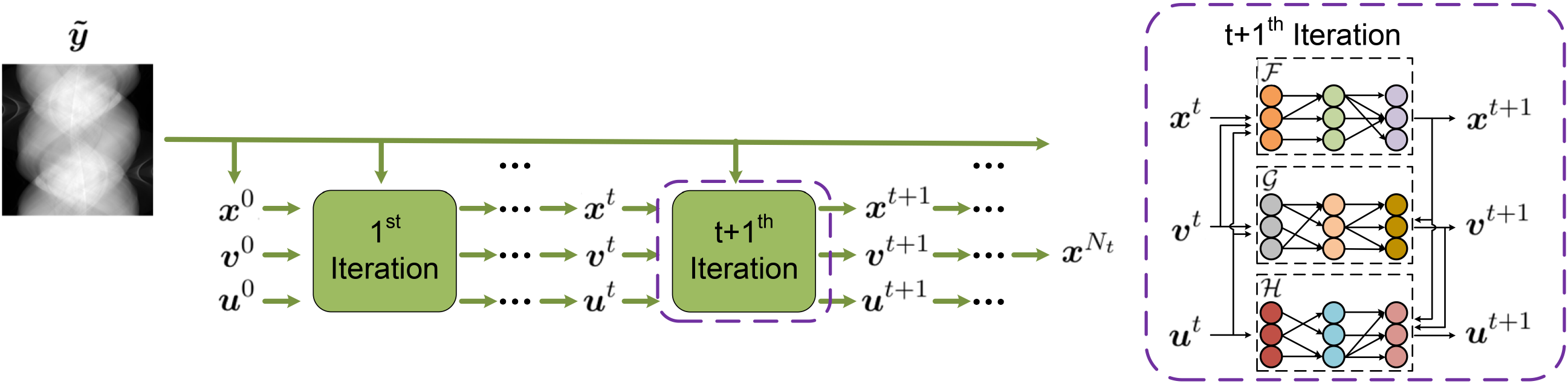}
	\caption{Workflow of an unrolled data-driven reconstruction process.}
	\label{fig:4}
\end{figure*}

\paragraph{Unrolling}
Unrolling is to expand the iterative optimization process into a finite number of stages and map them to a neural network~\cite{chen2018learn, adler2018learned,ding2021deep,zhang2020metainv, xiang2021fista, gilton2019neumann}.
As a general objective function, Eq.~\eqref{eq:12} can be extended using augmented Lagrangian method (ALM) as follows:
\begin{equation}
	\min_{\bm{x}, \bm{v}, \bm{u}} L(\bm{x}, \bm{v}, \bm{u}) = \frac{1}{2} (\bm{\tilde{y}}-A\bm{x})^T \Sigma^{-1} (\bm{\tilde{y}}-A\bm{x}) + \lambda R(\bm{v})
	+ \bm{u}^T(\bm{x}-\bm{v}) + \frac{\rho}{2}\|\bm{x}-\bm{v}\|_2^2.
	\label{eq:18}
\end{equation}
A general iterative optimization method ADMM can be formulated as
\begin{equation}
	\begin{split}
		\bm{x}^{t+1} &= \arg\min_{\bm{x}} \ L(\bm{x}, \bm{v}^t, \bm{u}^t),\\
		\bm{v}^{t+1} &= \arg\min_{\bm{v}} \ L(\bm{x}^{t+1}, \bm{v}, \bm{u}^t),\\
		\bm{u}^{t+1} &= \bm{u}^t + \rho(\bm{x}^{t+1} - \bm{v}^{t+1}).
	\end{split}
\label{eq:19}
\end{equation}
Each of these variables can be optimized using a corresponding algorithm. In an unrolling-based data-driven method, each optimization problem can be solved using a sub-network. When the total number of iterations is fixed, Eq.~\eqref{eq:19} can be realized as a neural network:
\begin{equation}
	\begin{split}
		\bm{x}^{t+1} &= \mathcal{F}(\bm{x}^t, \bm{v}^t, \bm{u}^t; \theta^t),\\
		\bm{v}^{t+1} &= \mathcal{G}(\bm{x}^{t+1}, \bm{v}^t, \bm{u}^t; \theta^t),\\
		\bm{u}^{t+1} &= \mathcal{H}(\bm{x}^{t+1}, \bm{v}^{t+1}, \bm{u}^t; \theta^t),
	\end{split}
\label{eq:20}
\end{equation}
where $\mathcal{F}$, $\mathcal{G}$ and $\mathcal{H}$ denote the three sub-networks, respectively.
Fig.~\ref{fig:4} shows a top-level of the workflow.

With different energy models and optimization algorithms, various network architectures were developed for unrolled data-driven image reconstruction.
In~\cite{chen2018learn}, the simplest gradient descent algorithm was unrolled into a neural network.
In~\cite{adler2018learned}, the primal-dual hybrid gradient (PDHG) algorithm was designed as a generalized unrolling technique.
In~\cite{xiang2021fista}, the momentum method commonly used for traditional optimization was adapted into a reconstruction network for  better performance with a limited number of iterations.
However, it is difficult to directly judge the performance of the network based on the performance of the unrolled optimization scheme,
and it remains an important topic how to unroll an optimization scheme and train it optimally.

\begin{figure*}[tbp]
	\centering
	\includegraphics[width=0.8\textwidth]{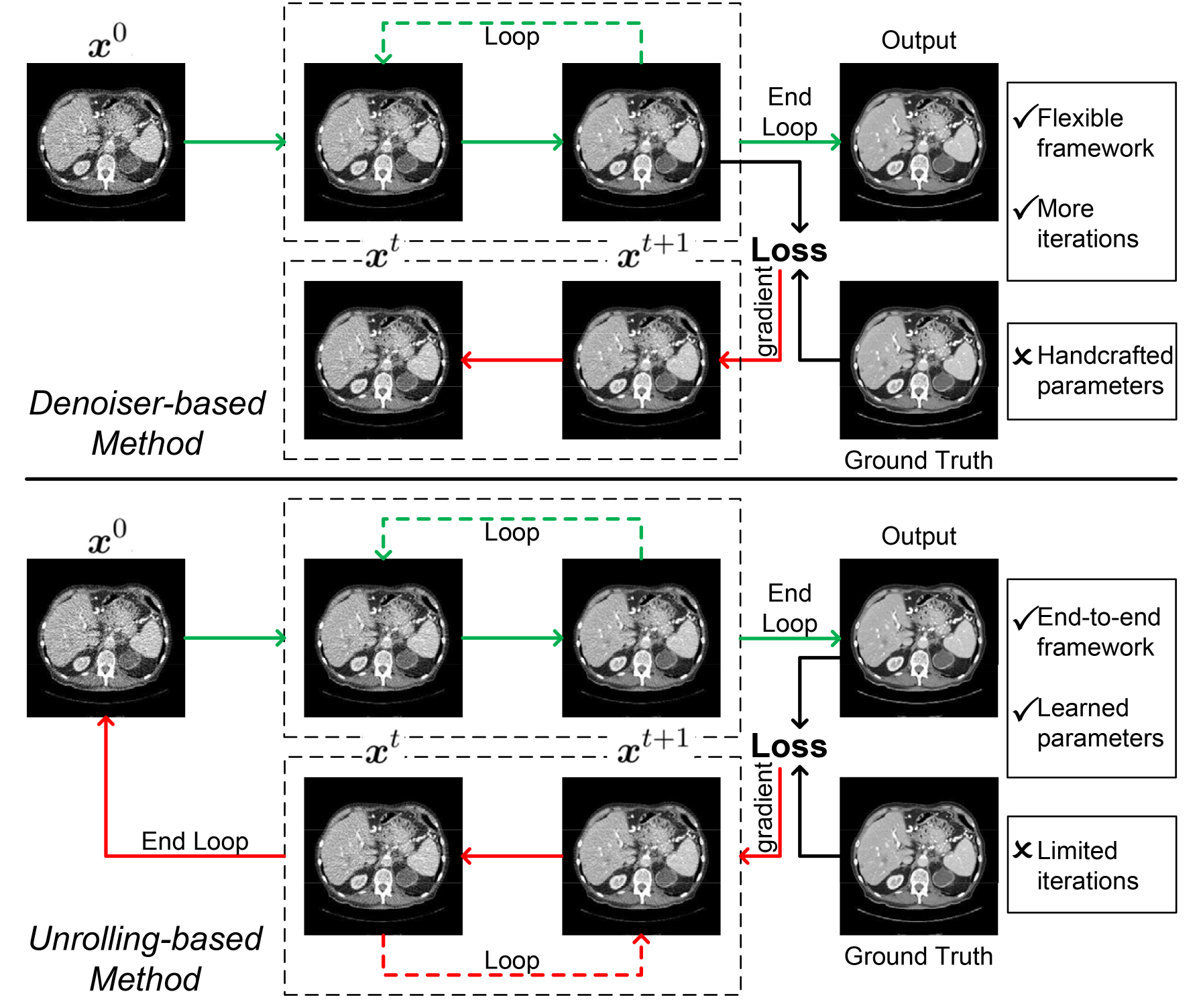}
	\caption{Forward and backward processes for of denoiser-based (top) and unrolling-based methods (bottom) respectively.}
	\label{fig:5}
\end{figure*}

The denoiser-based method is based on the traditional iterative algorithm, where the training and optimization are separated.
On the other hand, the unrolling-based method is an end-to-end procedure, where the optimization is incorporated into the training.
Fig.~\ref{fig:5} shows the forward and backward processes for denoiser-based and unrolling-based methods, where the green and red arrows represent forward and backward directions respectively.
As shown in Fig.~\ref{fig:5}, the forward data streams for the two methods are similar, but the backward data stream is end-to-end for the unrolling-based method; i.e., the
complete backward data stream is a back-projection of error signals from the output to the input.
While the denoiser-based method adopts a separate training strategy,
the unrolling-based method can be trained in a unified fashion, where all parameters, including the regularization parameters, can be obtained from training.
However, it is unavoidable that the model requires larger memory and thus limits the number of iterations for unrolling.
In many cases, the model performance is closely related to the number of iterations. 
In contrast, the denoiser-based method allows for more iterations and the trained denoiser can be embedded in different optimization schemes, making the denoiser-based method more flexible.
Nevertheless, the denoiser-based method still needs to set the parameters manually, which have a significant impact on the performance.
Hence, it is more important for the denoiser-based method to appropriately set parameters and be coupled with a method for adaptive parameter adjustment.

\subsection{Loss Functions}
The commonly used loss functions for LDCT imaging are mean square error (MSE) and mean absolute error (MAE).
To better remove noise and artifacts, total variation (TV) regularization, which performs well in compressed sensing methods for image denoising, is often used as an auxiliary loss~\cite{unal2022unsupervised}.
In~\cite{yang2018low}, the discriminator was used to make the denoised image have the same data distribution as that of clinical images.
Additionally, a model pre-trained for the classification task was used to extract features, and the perceptual loss was computed in the feature space.
The adversarial loss and perceptual loss can improve the visual performance and suppress the over-smoothness. However, the adversarial loss for generative adversarial networks may introduce erroneous  structures~\cite{ledig2017photo}. Similarly, the perceptual loss could generate checkerboard artifacts~\cite{sugawara2018super}, when the constraint is imposed on the feature space downsampled with maxpooling.
In~\cite{unal2022unsupervised}, the structural similarity index metric (SSIM) was introduced to promote structures closer to the ground truth.
Similarly, to protect edgeness in denoised images, the Sobel operator was applied to extract edges and keep the edge coherence~\cite{shan2019competitive}.
The identity loss is also relevant for image denoising tasks, which means that if a noise-free image is fed to the network then it should be dormant, i.e., the network output should be close to the clean input~\cite{kang2019cycle}.
To maintain the measurement consistency, the result of the network needs to be transformed into the projection domain to compute the MSE or MAE loss~\cite{unal2022unsupervised}.
Table~\ref{loss} summarizes these commonly used loss functions.

\begin{table}[tbp]
\centering
\small
\caption{Representative loss functions.}
\label{loss}
\begin{tabular}{lp{0.35\textwidth}<{\centering}p{0.32\textwidth}<{\centering}}
\toprule
\multicolumn{1}{c}{Reference} & \multicolumn{1}{c}{Pros} &
\multicolumn{1}{c}{Cons}\\
\midrule
MSE & Good denoising performance & Oversmoothing  \\
MAE & Good denoising performance & Oversmoothing \\
TV loss~\cite{unal2022unsupervised} &   Good denoising performance & Oversmoothing   \\
Adversarial loss~\cite{yang2018low} & Good visual effect & Erroneous structures \\
Perceptual loss~\cite{yang2018low}  &Good visual effect & Erroneous structures \\
SSIM loss~\cite{zhang2020metainv} &  Better structural protection  & \multicolumn{1}{c}{-} \\
Edge incoherence~\cite{shan2019competitive} & Better structural protection  & \multicolumn{1}{c}{-}    \\
Identity loss~\cite{kang2019cycle}  & More robust network  & \multicolumn{1}{c}{-} \\
Projection loss~\cite{unal2022unsupervised}  & 
Higher measurement consistency & Worse denoising performance\\ 
\bottomrule 
\end{tabular}
\end{table}

\section{Experimental Comparison}
In this section, we report our comparative study on the performance of some popular physics/model-based data-driven methods and different loss functions. 
This evaluation was performed with a unified code framework to ensure fairness as much as possible. 
All codes have been succinctly documented to help readers understand the models$\footnote{Codes are released at https://github.com/Deep-Imaging-Group/Physics-Model-Data-Driven-Review, related datasets and checkpoints can also be found on that page.}$.
For simplicity and fairness, the MSE loss function and AdamW optimizer were employed for all the methods when evaluating the models. 
And LPD was adopted as the backbone for the evaluation of different loss functions.
Training was performed in a naive way, without any trick. For fair comparison, all the models have been trained within 200 epochs, which is sufficient for convergence of all the methods. The penalty/regularization parameters of the models have been carefully tuned in our experiments to guarantee the optimal performance of each model on the relevant dataset.
After training, the optimal model for validation was taken as the final model and used for testing.
Of course, there are many factors that affect the performance of the neural network.
Therefore, the results in this paper are for reference only, which may not perfectly reflect the performance of these methods.

\subsection{Dataset}
The dataset used for our experiments is the public LDCT data from ``\textit{the 2016 NIH-AAPM-Mayo Clinic Low-Dose CT Grand Challenge}".
The dataset contains 2378 slices of 3 mm full-dose CT images from 10 patients.
In this study, 600 images from seven patients were randomly chosen as the training set, 
100 images from one patient were used as the validation set, 
and 200 images from the remaining two patients were the testing set.
The projection data were simulated with distance-driven method.
The geometry and volume were set according to the scanning parameters associated with the dataset.
The noise simulation was done using the algorithm in~\cite{yu2012development}.
The incident photon number for NDCT is the same as that provided in the dataset.
The incident photon number of LDCT was set to 20\% of that for NDCT.
The variance of the electronic noise was assumed to be 8.2 according to the recommendation in~\cite{yu2012development}.

\begin{figure*}[tbp]
	\centering
	\includegraphics[width=1.0\textwidth]{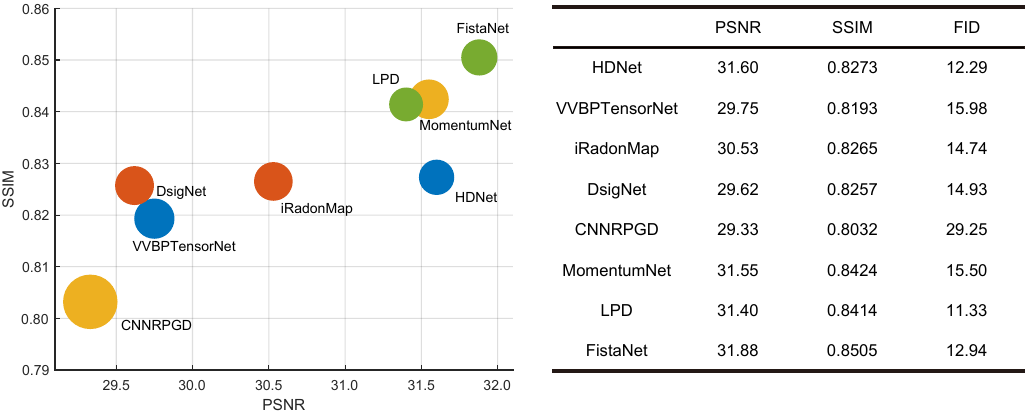}
	\caption{Quantitative results obtained using different methods on the whole testing set.}
	\label{fig:6}
\end{figure*}

\subsection{Model Study}
The training process for model evaluation was to minimize the MSE loss function with different kinds of the methods. The commonly used PSNR and SSIM metrics were adopted to quantify the performance of different denoising methods.
To evaluate the visual effect of the results, we introduced the Frechet inception distance (FID) score~\cite{NIPS2017_8a1d6947}.
A smaller FID score means a visual impression closer to the ground truth.
Fig.~\ref{fig:6} shows the means of PSNR, SSIM, and FID scores on the whole testing set. In Fig.~\ref{fig:6}, the 2D positions of the different methods are specified by the horizontal and vertical coordinates representing PSNR and SSIM of the results respectively, and the radii of the circles indicate the FID values of different methods.
It can be seen that the unrolling-based methods have more robust performance. FistaNet and LPD are in favorable spots. The denoiser-based methods also have outstanding performance, especially MomentumNet based on an iteration-dependent denoiser. The comparison between MomentumNet and CNNRPGD shows that iteration-dependent denoiser has clearly better performance. However, the training of an iteration-dependent denoiser is more 
complicated and time-consuming. The training time of MomentumNet for 200 epochs is more than 5 days, which is much longer than that needed by CNNRPGD. Additionally, the denoiser-based methods need manually setting regularization parameters, which often has a greater impact on the performance than the network architecture and requires a major fine-tuning effort.
HDNet delivers the best performance among the physics-based methods, which proves that the simple FBP transform is effective for dual-domain-based reconstruction.
For the learned transform-based method, since the FC layers is of a large scale, the training process is relatively difficult, compromising the stability of reconstruction results.

\begin{table}[tbp]
\centering
\small
\caption{Computational costs of the compared methods.}
\label{cost}
\begin{tabular}{cccc}
\toprule
Method        & Training time & Testing time & Number of Parameters \\
\midrule
HDNet         & 11.67 h       & 0.16 s       & 75.3 M                \\
VVBPTensorNet & 9.77 h        & 0.19 s       & 0.47 M                \\
\midrule
iRadonMap     & 11.30 h       & 0.12 s       & 270 M                 \\
DSigNet       & 9.58 h        & 0.27 s       & 19.2 M                \\
\midrule
CNNRPGD       & 3.10 h        & 3.22 s       & 34.6 M                \\
MomentumNet   & 122.50 h      & 1.16 s       & 7.5 M                 \\
\midrule
LPD           & 7.95 h        & 0.20 s        & 0.25 M               \\
FistaNet      & 7.83 h        & 0.21 s       & 0.78 M                \\
\bottomrule
\end{tabular}
\end{table}

Table~\ref{cost} shows the computational time of the compared methods. It can be seen that most methods can complete the reconstruction in a short time, which is beneficial for clinical applications. Given a large number of iterations, the computational time of the denoiser-based iterative methods are much greater than that of other methods.

Unlike the unrolling-based methods which are end-to-end networks, the denoiser-based methods are implemented in the iterative framework. Therefore, it is important to study their convergence properties. In~\cite{gupta2018cnn} and~\cite{chun2020momentum}, the authors proved that the denoiser-based iterative methods can converge. Furthermore, 
even if the denoiser is applied to other iterative optimization schemes with a good convergence property, they should converge similarly, which demand a more rigorous justification in the future.

\begin{table}[tbp]
\centering
\small
\caption{Loss functions used for experimental comparison.}
\label{loss_combination}
\begin{tabular}{lcccccccccc}
\toprule
&c&d&e&f&g&h&i&j&k&l\\
\midrule
MSE& \checkmark&& \checkmark& \checkmark&& \checkmark& \checkmark& \checkmark& &\\
MAE&& \checkmark&&&&&&&&\\
TV loss&&&&&&&&&& \checkmark\\
Adversarial loss&&& \checkmark&& \checkmark&&& \checkmark& \checkmark&\\
Perceptual loss&&&& \checkmark& \checkmark&&&&&\\
SSIM loss&&&&&& \checkmark&&&&\\
Edge incoherence&&&&&&& \checkmark& \checkmark&&\\
Identity loss&&&&&&&&& \checkmark&\\
Projection loss&&&&&&&&&& \checkmark\\
\midrule
supervised learning&\checkmark&\checkmark&\checkmark&\checkmark&\checkmark&\checkmark&\checkmark&\checkmark&&\\
unsupervised learning&&&&&&&&&\checkmark&\checkmark\\
\bottomrule 
\end{tabular}
\end{table}

\subsection{Loss Function Study}
\begin{figure*}[tbp]
	\centering
	\includegraphics[width=1.0\textwidth]{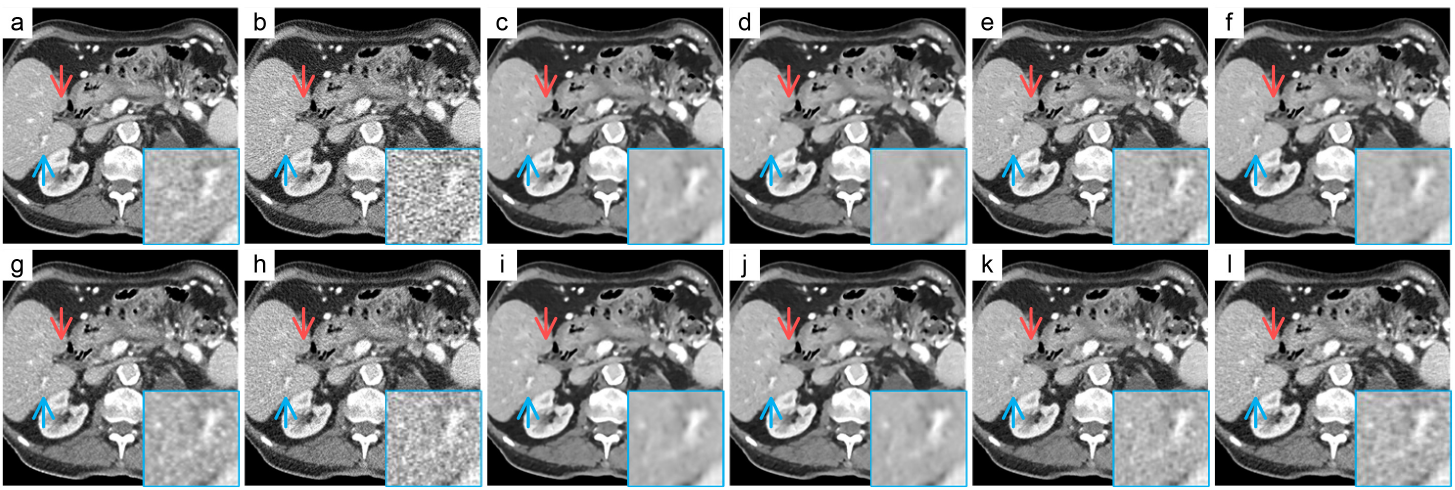}
	\caption{Results obtained using LPD with different combinations of loss functions. (a) Normal-dose, (b) Low-dose, (c)-(l) the reconstructions with the combinations of loss functions shown in Table~\ref{loss_combination}. The display window is [-160, 240] HU.}
	\label{fig:7}
\end{figure*}

To evaluate the effects of loss functions, we have combined the loss functions in various ways and applied each representative combination to a unified LPD model. Table~\ref{loss_combination} shows these combinations of the loss functions. The corresponding reconstructions of an abdominal slice are shown in Fig.~\ref{fig:7}. Note that the weights of combinations have been fine-tuned experimentally for optimal visibility. In Fig.~\ref{fig:7}, the LPDs trained with different loss functions can all keep the key information on the metastases indicated by the red arrows. The area indicated by the blue arrows is enlarged for better visualization. Based on the same network architecture, while the restored information of different results is basically the same, the main difference among them can be still visually appreciated. MSE and MAE have evident an over-smoothing effect. The adversarial and perceptual losses can effectively improve the visual impression, 
giving the reconstructed textures similar to the ground truth. With the help of the adversarial and TV losses, the network can achieve satisfactory results via unsupervised learning.

\section{Discussions}
Physics/model-based data-driven methods have received increasing attention in the tomographic imaging field because they incorporate the CT physics or models into the neural networks synergistically, resulting in superior imaging performance. 
With rapid development over the past years, researchers have proposed a number of models based on physics/models from different angles. 
Although they are promising, these models still need further improvements. 
We believe that the following issues are worth further investigation. 
The first issue is the generalizability of learned transform-based data-driven methods. 
Training the networks separately for each imaging geometry is an unaffordable cost in clinical applications. 
Therefore, a major problem with these methods is to make a trained model applicable to multiple geometries and volumes. 
Interpolation can help match sizes of input data, required by a reconstruction network.
Furthermore, a deep learning method can be a good solution to convert projection data from a source geometry to a target geometry.
The second topic is the parametric setting for the denoiser-based data-driven methods. 
Currently, this kind of method requires handcrafted setting, which limits its generalization to different datasets. 
The introduction of adaptive parameters or learned parameters is worthy of attention. Reinforcement learning could be another option to automatically select hyper-parameters.
The above are of our specific interest for physics/model-based data-driven methods for LDCT.
From a larger perspective, the tomographic imaging field has other open topics and challenges, which are also closely related to LDCT.

\paragraph{Transformer}
Transformer is an emerging technology of deep learning.
It has shown great potential in various areas~\cite{liu2021swin}.
In the denoising task, a transformer directs attention to various important features, resulting in adaptive denoising based on image content and features. Coupled with the transformer, physics/model-based data-driven methods will have more design routes.
It is predictable that transformers will further improve the performance of physics/model-based data-driven methods.

\paragraph{Self-supervised learning}
Paired training data has always been a conundrum plaguing data-driven tomography.
The mainstream method is now unsupervised~\cite{kang2019cycle} and self-supervised learning~\cite{kim2021noise2score}, which does not require paired/labeled data.
Self-supervised training treats the input as the target in appropriate ways to calculate losses, and performs denoising according to the statistical characteristics of underlying data.
Clearly, a combination of self-supervised training and physics/model-based data-driven methods can help us meet the challenge of LDCT in clinical applications.

\paragraph{Task-driven tomography}
Tomographic imaging is always a service for diagnosis and intervention. Thus, reconstructed images are often processed or analyzed before being clinically useful.
To optimize the whole workflow, we can take the downstream image analysis tasks into account to improve the performance of reconstruction network in a task-specific fashion.
The physics/model-based task/data-driven method can be designed with shared feature layers linked to task loss functions.
A deep tomographic imaging network incorporated with a task-driven technique can reconstruct results that are more suitable for the intended task in terms of diagnostic performance.

\paragraph{Domain generalization}
Deep learning-based tomographic imaging may suffer from a domain heterogeneity problem from different distributions of training data, which originate from different scanners, populations, tasks, settings, and so on~\cite{xia2021ct}.
Existing tomographic imaging methods could generalize poorly on datasets in shifted domains, especially unseen ones.
Domain generalization is to learn a model from one or several different but related domains, which has attracted increasing attention~\cite{wang2022generalizing}.
This is a promising direction to address the data domain heterogeneity and  advance the clinical translation of deep tomographic imaging methods.

\paragraph{Image quality assessment}
At present, the main means to evaluate reconstructed image quality is still mostly the popular quantitative metrics. 
But in many cases the classic quantitative evaluation is not consistent with the visual effects and clinical utilities. 
Especially, the way to evaluate medical images is very different from that of natural images. 
Therefore, it is currently an open problem to have a set of metrics suitable to evaluate the diagnostic performance of tomographic imaging. 
For natural image processing, there are neural networks reported for image quality assessment (IQA)~\cite{wang2004image}, which suggest new solutions for medical image quality evaluation. 
Ideally, DL-based IQA should not only judge the reconstruction quality and diagnostic performance but also help tomographic imaging in the form of loss functions.
It is expected that more DL-based IQA methods will be developed for medical imaging, and eventually can perform advanced numerical observer studies as well as human reader studies. 


\section{Conclusion}
In this paper, we have systematically reviewed the physics/model-based data-driven methods for LDCT.
In important clinical applications of LDCT imaging, DL-based methods bring major gains in image quality and diagnostic performance and are undoubtedly becoming the mainstream of LDCT imaging research and translation.
In the next few years, our efforts would cover dataset enrichment, network adaption, and clinical evaluation, as well as methodological innovation and theoretical investigation.
From a larger perspective, DL-based tomographic imaging is only in its infancy. It offers many problems to solve for numerous healthcare benefits 
and opens a new era of AI-empowered medicine.


\small{
\bibliographystyle{IEEEtran} 
\bibliography{reference.bib}
}
\end{document}